\magnification=1200
\parskip 10pt plus 5pt
\parindent 14pt
\baselineskip=18pt
\pageno=0
\footline={\ifnum \pageno <1 \else \hss \folio \hss \fi}
\line{\hfil{DAMTP-R/96/7~}}
\line{\hfil{gr-qc/9608025~~~~~}}
\line{\hfil{(Updated~version)}}
\topglue 1in
\centerline{\bf SYMMETRIC VARIATIONS OF THE METRIC AND}
\vskip 1pt
\centerline{\bf EXTREMA OF THE ACTION
FOR PURE GRAVITY}
\vskip .65in
\centerline{Simon Davis}
\vskip .4in
\centerline{Department of Applied Mathematics and Theoretical Physics}
\vskip 1pt
\centerline{University of Cambridge}
\vskip 1pt
\centerline{Silver Street, Cambridge CB3 9EW, England}
\vskip .5in
\noindent{\bf Abstract}.  Symmetries of generalized gravitational actions,
yielding field equations which typically involve at most second-order 
derivatives of the metric, are considered.  The field equations for several
different higher-derivative theories in the first-order formalism are
derived, and variations of a generic set of higher-order curvature terms
appearing in string effective actions are studied.  It is shown that there 
often exists a particular set of solutions to the field equations of pure 
gravity theories, consisting of different combinations of curvature
tensors, which satisfies the vacuum equations with cosmological constant.
Implications of generalized symmetries of the field equations derived from
the superstring effective action for the cosmological constant problem are
discussed.
\vfill
\eject

\noindent{\bf 1. Introduction}

Recent interest in generalized gravitational actions [1][2][3] has arisen 
because of the appearance of higher-derivative curvature terms in superstring 
effective
actions and the extension of the Hilbert-Palatini formalism using different 
choices of canonical variables for the phase space of general relativity 
[4][5][6].  These generalizations  
have been studied with the understanding that they will allow a more flexible 
interpretation of gravity at the level of the classical action and its 
quantization, while still retaining the framework of the Einstein 
gravitational field equations in the classical limit.  One of the 
consequences of 
this approach is that additional symmetries, besides
diffeomorphism invariance, can be incorporated, especially in 
the context of superstring theory.  

These symmetries can be used to derive new results about the cosmological 
constant.  Although the cosmological constant problem, as it is presently 
stated [7][8], concerns the contrast between the sum of zero-point energy 
densities of fields in elementary-particle Lagrangians at the symmetry breaking
scales and the observed value of the vacuum energy density, the {\it classical}
version of the problem, initially arising in the replacement of 
de Sitter space by the Friedman-Robertson-Walker universe [7], can be 
extended to fitting all realistic cosmological models with metrics 
corresponding to $\Lambda~\simeq~0$.  This version of the problem, 
relevant at macroscopic
scales, cannot be solved using classical techniques applied to the 
standard Einstein-Hilbert action, and quantum fluctuations of the
metric at Planck scales do not necessarily determine the cosmological 
structure at large scales.

The classical version of the cosmological constant problem can be 
studied within the context of a generalized gravitational action.  
Generalizations of the derivation of the field equations from the 
Einstein-Hilbert action shall be considered from several different 
viewpoints.  First, in a preliminary study of variations of 
the standard action with respect to both symmetric and non-symmetric metrics, 
the distinctions between the two cases will be noted
in the variation of ${\sqrt {-g}}$ with respect to $g_{\mu\nu}$
and adapted to the tetrad formalism. Second, the difference in the functional
derivative with respect to $g^{\mu\nu}$ suggests the 
consideration of variations of actions which include arbitrary coefficients
multiplying $g_{\mu\nu}$.  A suitable averaging of such integrals, initially 
written in a different form from the standard Einstein-Hilbert action, is used 
to derive a set of field equations which limit the size of the cosmological 
constant when there are gravitational perturbations.  Third, 
higher-derivative actions, particularly those which arise in string theory, 
are used to place the cosmological constant in a different setting.  
For example, it is recalled in $\S 5$ how the vacuum Einstein field equations 
with non-zero cosmological constant can be derived from an action containing 
higher-order curvature terms but no cosmological constant term.  It is also
noted how the additional symmetries associated with string theory have been 
used to shift the value of the cosmological constant.  The concept of the most
general class of symmetries of the gravitational field 
equations is developed further in $\S 7$ to demonstrate the special place of 
the $\Lambda=0$ sector in the space of solutions to the field equations.  
Scale transformations are then used to provide a 
resolution to the classical version of the cosmological constant 
problem.  Finally, the results 
in this paper are discussed in the context of other attempts to resolve the 
quantum version of the cosmological constant problem. 

\noindent{\bf 2. Symmetric and Non-Symmetric Metric Variations of the Action}

As the field equations derived from the action
$${1\over {16 \pi G}} \int~d^4 x \sqrt{-g(x)}~(R~-2 \Lambda)~+~
\int~d^4 x \sqrt{-g(x)} L_{matter}
\eqno(1)$$
depend on the variation of the determinant $g$, the expansion into minors
$$det(g_{\mu\nu})~=~\sum_{\nu}~(-1)^{\mu +\nu}~[M(g)]_{\mu\nu} g_{\mu\nu}
\eqno(2)$$
with $[M(g)]_{\mu\nu}$ being the determinant of the $3\times 3$ matrix 
excluding
the $\mu^{th}$ column and the $\nu^{th}$ row, can be used to obtain the term
proportional to $g_{\mu\nu}$.  When $\mu \ne \nu$, $[M(g)]_{\mu\nu}$ includes
the entry $g_{\nu\mu}$.  If the symmetry of the metric is not imposed, then
$g_{\mu\nu}$ and $g_{\nu\mu}$ may be regarded as independent variables
[9][10][11].
Under these conditions, $[M(g)]_{\mu\nu}$ is independent of $g_{\mu\nu}$ and
$${{\delta g}\over {\delta g_{\mu\nu}}}~=~(-1)^{\mu+\nu}[M(g)]_{\mu\nu}
\eqno(3)$$
Since $g\cdot g^{\mu\nu}~=~(-1)^{\mu+\nu}[M(g)]_{\nu\mu}$, equality of
${{\delta g}\over {\delta g_{\mu\nu}}}$ with $g\cdot g^{\mu\nu}$ can be
obtained when $[M(g)]$ is a symmetric matrix.

An alternative expression for the determinant utilizes the splitting of the
new metric $g_{\mu\nu}$ into the original metric 
${g^{\raise2pt\hbox{$\scriptstyle{\!\!\! \circ}$}}}_{\mu\nu}$ and the
variation $\delta g_{\mu\nu}$.
$$det(g_{\mu\nu})~=~ 
det({g^{\raise2pt\hbox
{$\scriptstyle{\!\!\! \circ}$}}}_{\mu\nu}+\delta g_{\mu\nu})~=~
\sum_\nu~(-1)^{\mu+\nu}~[M(g^{\raise2pt\hbox{$\scriptstyle{\!\!\! \circ}$}}
 +\delta g)]_{\mu\nu}
({g^{\raise2pt\hbox{$\scriptstyle{\!\!\! \circ}$}}}_{\mu\nu}~+~\delta 
g_{\mu\nu})
\eqno(4)$$
As
$$[M({g^{\raise2pt\hbox{$\scriptstyle{\!\!\! \circ}$}}}+
\delta g)]_{\mu\nu}~=~
[M(g^{\raise2pt\hbox{$\scriptstyle{\!\!\! \circ}$}})]_{\mu\nu}
~+~\sum_{\alpha,\beta} 
~{{\partial [M(g)]_{\mu\nu}}\over {\partial g_{\alpha\beta}}}
\Big|_{M(g)=M(g^{\raise2pt\hbox{$\scriptstyle{\!\!\! \circ}$}})}
~\delta g_{\alpha\beta}~+~...
\eqno(5)$$ 
it follows that
$${{\delta g}\over {\delta g_{\mu\nu}}}\Big|_{g_{\mu\nu}=
{g^{\raise2pt\hbox{$\scriptstyle{\!\!\! \circ}$}}}_{\mu\nu}}~=~
(-1)^{\mu + \nu}~
[M(g^{\raise2pt\hbox{$\scriptstyle{\!\!\! \circ}$}})]_{\mu\nu}
\eqno(6)$$
when $\delta g_{\mu\nu}$ is non-symmetric, and
$${{\delta g}\over {\delta g_{\mu\nu}}}\Big|_{g_{\mu\nu}=
{g^{\raise2pt\hbox{$\scriptstyle{\!\!\! \circ}$}}}_{\mu\nu}}~=~
(-1)^{\mu +\nu}~
[M(g^{\raise2pt\hbox{$\scriptstyle{\!\!\! \circ}$}})]_{\mu\nu}~+~
\sum_\nu~(-1)^{\alpha + \nu}
{{\partial [M(g)]_{\alpha \nu}}\over {\partial g_{\nu\mu}}}
\Big|_{M(g)=M(g^{\raise2pt\hbox{$\scriptstyle{\!\!\! \circ}$}})}~
{g^{\raise2pt\hbox{$\scriptstyle{\!\!\! \circ}$}}}_{\alpha\nu}
\eqno(7)$$
when $\delta g_{\mu\nu}$ is symmetric and
${g^{\raise2pt\hbox{$\scriptstyle{\!\!\! \circ}$}}}_{\mu\nu}~=~
{g^{\raise2pt\hbox{$\scriptstyle{\!\!\! \circ}$}}}_{\nu\mu}$.

The symmetry of the background metric is sufficient for the equality of
${{\delta g}\over {\delta g_{\mu\nu}}}\Big|_{g_{\mu\nu}=
g^{\raise2pt\hbox{$\scriptstyle{\!\!\! \circ}$}}_{\mu\nu}}$
and $g^{\raise 2pt\hbox{$\scriptstyle{\!\!\! \circ}$}}
g^{\raise2pt\hbox{$\scriptstyle{\!\!\! \circ}$}{\mu\nu}}$ when $\delta 
g_{\mu\nu}$ is non-symmetric.  The
equality no longer holds for general metrics, for which $[M(g)]_{\mu\nu}
\ne [M(g)]_{\nu\mu}$.  Thus, for non-symmetric variations $\delta g_{\mu\nu}$,
the term proportional to the metric arises in the field equations only for the 
backgrounds with symmetric metric 
${g^{\raise2pt\hbox{$\scriptstyle{\!\!\! \circ}$}}}_{\mu\nu}$.

In the first-order formalism, variations of the gravitational action
with respect to both $g_{\mu\nu}$ and $\Gamma^\rho{}_{\mu\nu}$ are considered,
so that the connection is shown to be Levi-Civita and the field
equations are derived for the metric.
The restriction to the symmetric background is also required for the variation
of the Ricci scalar. Defining the curvature tensor by parallel transport 
around 
the integral curves of two independent vector fields, it can be shown that
$$(\delta {\Gamma^{\lambda}}_{\mu\lambda})_{; \nu}~-~(\delta 
{\Gamma^{\lambda}}_{\mu\nu})_{; \lambda}~=~\delta R_{\mu\nu}~+~2{T^{\rho}}_{\nu
\lambda} \delta {\Gamma^{\lambda}}_{\mu\rho}
\eqno(8)$$
where ${T^{\rho}}_{\nu\lambda}~=~{\Gamma^{\rho}}_{[\nu\lambda]}$ is the torsion
tensor.  Since
$${T^{\rho}}_{\nu\lambda}~=~{1\over 2} g^{\rho\tau}(\delta g_{[\tau\lambda], 
\nu}
~+~\delta
g_{[\nu\tau], \lambda}~-~\delta g_{[\nu \lambda], \tau})
\eqno(9)$$
it will vanish for the symmetric background and
$$\delta R_{\mu\nu}
\Big|_{g_{\mu\nu}={g^{\raise2pt
\hbox{$\scriptstyle{\!\!\! \circ}$}}}_{\mu\nu}}~=~
(\delta {\Gamma^{\lambda}}_{\mu\lambda})_{; \nu}
\Big|_{g_{\mu\nu}={g^{\raise2pt\hbox{$\scriptstyle{\!\!\!
 \circ}$}}}_{\mu\nu}}~-~
(\delta 
{\Gamma^{\lambda}}_{\mu\nu})_{; \lambda}
\Big|_{g_{\mu\nu}={g^{\raise2pt\hbox{$\scriptstyle{\!\!\! \circ}$}}}_{\mu\nu}}
\eqno(10)$$ 
$$\sqrt{-g} g^{\mu\nu} \delta R_{\mu\nu} 
\Big|_{g_{\mu\nu}=g^{\raise2pt\hbox{$\scriptstyle{\!\!\! \circ}$}}_{\mu\nu}}~=~
{{\partial}\over {\partial x^{\nu}}}
[{\sqrt {-g^{\raise2pt\hbox{$\scriptstyle{\!\!\! \circ}$}} }} 
(g^{\raise2pt\hbox{$\scriptstyle{\!\!\! \circ}$}{\mu\nu}} \delta 
{\Gamma^{\raise4pt
\hbox{$\scriptstyle{\!\!\! \circ}$}{\lambda}}}{}_{\mu\lambda}~-~
g^{\raise2pt\hbox{$\scriptstyle{\!\!\! \circ}$}{\mu\lambda}}~\delta 
{\Gamma^{\raise4pt\hbox{$\scriptstyle{\!\!\! \circ}$}{\nu}}}{}_{\mu\lambda})]
\eqno(11)$$
As the integral can be transformed to one associated with a surface integral, 
which is 
zero when
$\delta g_{\mu\nu}$ and its first derivatives vanish at
infinity, the variation of the action becomes
$${1\over {16 \pi G}} \delta \int d^4 x \sqrt {-g(x)} (R - 2 \Lambda)
\Big|_{g_{\mu\nu}= 
{g^{\raise2pt\hbox{$\scriptstyle{\!\!\! \circ}$}}}_{\mu\nu}}=
{1\over {16 \pi G}}\int
d^4 x {\sqrt{-g^{\raise2pt\hbox{$\scriptstyle{\!\!\! \circ}$}} }} 
({R^{\raise4pt\hbox{$\scriptstyle{\!\!\! \circ}$}}}_{\mu\nu}~-~{1\over 2}
R^{\raise4pt\hbox{$\scriptstyle{\!\!\! \circ}$}}
g_{\mu\nu}~+~\Lambda 
{g^{\raise2pt\hbox{$\scriptstyle{\!\!\! \circ}$}}}_{\mu\nu}) 
\delta 
g^{\mu\nu}
\eqno(12)$$
for non-symmetric variations $\delta g^{\mu\nu}$.  The restriction to the
background metric 
${g^{\raise2pt\hbox{$\scriptstyle{\!\!\! \circ}$}}}_{\mu \nu}$ implies that 
only the tangent vector in the
directions of the non-symmetric variations $\delta g^{\mu \nu}$ vanishes at 
the solutions of the field equations
$${R^{\raise4pt\hbox{$\scriptstyle{\!\!\! \circ}$}}}_{\mu\nu}~-~{1\over 2}
~R^{\raise4pt\hbox{$\scriptstyle{\!\!\! \circ}$}} 
{g^{\raise2pt\hbox{$\scriptstyle{\!\!\! \circ}$}}}_{\mu\nu}~+~\Lambda  
{g^{\raise2pt\hbox{$\scriptstyle{\!\!\! \circ}$}}}_{\mu\nu}~=~0
\eqno(13)$$

For symmetric variations $\delta g_{\mu\nu}$, the torsion tensor in equation
(9) is zero, but the extra term in equation (7) contributes to the
variation of the action as

$$\eqalign{{{\partial [M(g)]_{\alpha\nu}}\over {\partial g_{\nu\mu}}}~=~
(-1)^{\mu +\nu}~[M^\prime(g)]_{\alpha\nu,\nu\mu}~~~~~~~~~~when~&
\alpha < \nu <\mu ~or~ \alpha > \nu >\mu
\cr
(-1)^{\mu + \nu +1} [M^\prime(g)]_{\alpha \nu, \nu\mu}~~~~~~~~~~
when~&\alpha < \nu, ~\mu < \nu~or ~ \alpha > \nu,~\mu > \nu
\cr}
\eqno(14)$$
where $[M^{\prime}(g)]_{\alpha\nu, \nu\mu}$ is the determinant of the 
$2\times 2$ matrix excluding
 the $\alpha^{th}$ and $\nu^{th}$ rows and the $\mu^{th}$
and $\nu^{th}$ columns.  Since
$$\eqalign{\sum_{{{{\nu\atop {\alpha < \nu <\mu}} \atop {or}} \atop 
{\alpha > \nu > 
\mu}}}~
(-1)^{\alpha + \nu}~
{g^{\raise2pt\hbox{$\scriptstyle{\!\!\! \circ}$}}}_{\alpha\nu}
[M^\prime(g^{\raise2pt\hbox{$\scriptstyle{\!\!\! \circ}$}})]_{\alpha \nu, 
\nu\mu}
~=&~\sum_{{{{\nu\atop {\alpha < \nu, \mu < \nu}} \atop {or}} \atop {\alpha > 
\nu, \mu > \mu}}}~
(-1)^{\alpha + \nu + 1} 
{g^{\raise2pt\hbox{$\scriptstyle{\!\!\! \circ}$}}}_{\alpha\nu} 
[M^\prime(g^{\raise2pt\hbox{$\scriptstyle{\!\!\! \circ}$}})]_{\alpha \nu, 
\nu\mu}
\cr
=&~[M(g^{\raise2pt\hbox{$\scriptstyle{\!\!\! \circ}$}})]_{\nu\mu}
\cr}
\eqno(15)$$
for symmetric variations $\delta g_{\mu \nu}$,
$${{\delta g}\over {\delta g_{\mu\nu}}}\Big|_{g_{\mu\nu}~=~
{g^{\raise2pt\hbox{$\scriptstyle{\!\!\! \circ}$}}}_{\mu\nu}}
~=~(-1)^{\mu + \nu}~
([M(g^{\raise2pt\hbox{$\scriptstyle{\!\!\! \circ}$}})]_{\mu\nu}~+~
[M(g^{\raise2pt\hbox{$\scriptstyle{\!\!\! \circ}$}})]_{\nu\mu})~=
~g^{\raise2pt\hbox{$\scriptstyle{\!\!\! \circ}$}} \cdot
(g^{\raise2pt\hbox{$\scriptstyle{\!\!\! \circ}$} {\mu\nu}}~+~
g^{\raise2pt\hbox{$\scriptstyle{\!\!\! \circ}$} {\nu \mu}})
\eqno(16)$$
so that $\delta g~=~g\cdot g^{\mu \nu} \delta g_{\mu \nu}$, and the variation
of the gravitational action is again given by equation (12).
While the field equations remain unchanged, one may note the factor of two
implicit in equation (16).  This factor has been omitted sometimes in previous
derivations of the field equations [12][13], presumably because it is 
absorbed in the summation over the indices $\mu$ and $\nu$.  It can be 
confirmed 
for Euclidean metrics by the calculation of the variation
of the determinant represented as $exp[tr~ln(\delta_{\mu \nu}~+~h_{\mu \nu})]$.

This property of the variation $\delta g$ is also evident when components other
than those obtained by interchange of the indices are dependent.
When
$$g_{\mu\nu}~=~\left(\matrix{-\alpha&\epsilon& \epsilon& \epsilon
                              \cr
                             \epsilon& \alpha&\epsilon& \epsilon
                              \cr
                              \epsilon&\epsilon& \alpha&\epsilon
                              \cr
                              \epsilon& \epsilon& \epsilon& \alpha
                               \cr}
                              \right)
\eqno(17)$$
${{\delta g}\over {\delta g_{00}}}\bigl\vert_{\alpha=1,\epsilon=0}~=~
{{\partial g}\over {\partial \alpha}}~=~4$ while
$g\cdot g^{00}\bigl\vert_{\alpha=1, \epsilon=0}~=~1$, but
$$g (g^{00}\delta g_{00}~+~g^{11} \delta g_{11}~+~g^{22} \delta g_{22}
~+~g^{33} \delta g_{33})\bigl\vert_{\alpha=1, \epsilon=0}~=~4~\delta\alpha
\eqno(18)$$
Similarly, if
$$g_{\mu\nu}~=~\left(\matrix{\alpha_0&\epsilon&\epsilon&\epsilon
                               \cr
                              \epsilon&\alpha_1& \epsilon&\epsilon
                               \cr
                               \epsilon&\epsilon&\alpha_2&\epsilon
                                \cr
                                \epsilon&\epsilon&\epsilon&\alpha_3
                                \cr}
                                 \right)
\eqno(19)$$

$$\eqalign{ {{\delta g}\over {\delta g_{01}}}~=~{{\partial g}\over {\partial 
\epsilon}}
~=~\alpha_0(6\epsilon^2&-2(\alpha_1+\alpha_2+\alpha_3)\epsilon)
\cr
~&-~2\epsilon(\alpha_1\alpha_2+\alpha_1\alpha_3+\alpha_2\alpha_3)
~+~6(\alpha_1+\alpha_2+\alpha_3)\epsilon^2-12 \epsilon^3
\cr}
\eqno(20)$$
while
$g\cdot g^{01}~=~-\epsilon(\alpha_2-\epsilon)(\alpha_3-\epsilon)$ but

$$\eqalign{2(g \cdot g^{01}\delta g_{01}+g\cdot g^{02} \delta g_{02}+&g
\cdot g^{03}
\delta g_{03}+g\cdot g^{12} \delta g_{12}+g\cdot g^{13}\delta g_{13}
+g \cdot g^{23} \delta g_{23})
\cr
~=~[\alpha_0(6\epsilon^2-&2(\alpha_1+\alpha_2+
\alpha_3)\epsilon)-2\epsilon(\alpha_1\alpha_2+\alpha_1\alpha_3+\alpha_2
\alpha_3)
\cr
&~~~~~~~~~~~~~~~~+6(\alpha_1+\alpha_2+\alpha_3)\epsilon^2
-12\epsilon^3]~\delta\epsilon
\cr}
\eqno(21)$$
consistent with the equality of $\delta g$ and $g\cdot g^{\mu\nu}\delta 
g_{\mu\nu}$.
    
In an alternative derivation of the field equations using the tetrad 
formulation
of the action
$$S~=~{1\over 2}~\int_M~d^4x~\epsilon^{\mu \nu \rho\sigma}\epsilon_{abcd}
\left[~e_\mu^a e_\nu^b {R_{\rho\sigma}}^{cd}(\omega)~-~{\Lambda \over 3}
e_\mu^a e_\nu^b e_\rho^c e_\sigma^d \right]
\eqno(22)$$
one uses the variation of the determinant of the tetrad
$$\eqalign{{{\delta(det~e)}\over {\delta~e_\mu{}^a}}~=&~
{(e^{-1})^{[a]}}_{[\mu]}~(det~e)
\cr
[a]~=&~space-time~index~in~the~same~numerical~order~as~a
\cr
[\mu]~=&~tangent~space~index~in~the~same~numerical~order~as~\mu
\cr}
\eqno(23)$$
which follows from the expansion of minors as it did for the non-symmetric 
metric.  This nomenclature is being used to clarify the connection between the
the tetrad and its inverse.

The inverse tetrad satisfies
$(e^{-1})^{[b]}{}
_{[\mu]}~{e_{[b]}}^a~=~\delta_{[\mu]}{}^a$.
Defining the matrix $(e^{-1})$ to be $(e^\mu{}_a)$, the relation satisfied 
by the inverse tetrad becomes ${e^\mu}_a {e_\mu}^b~=~{\delta_a}^b$ after 
relabelling of the indices.  Equation (23) is equivalent, therefore, to the 
standard relation
${{\delta(det~e)}\over {\delta {e_\mu}^a}}~=~{e^\mu}_a~(det~e)$ after 
interpretation of the matrix $(e^\mu{}_a)$ as the inverse of the tetrad.  
This definition is required because of the non-symmetry of the
tetrad, $e_\mu{}^a ~\not=~e_{[a]}{}^{[\mu]}$, as a matrix with the rows 
labelled by space-time indices and the columns labelled by tangent-space 
indices.

\noindent {\bf 3.  Generalized Gravitational Actions in Two Dimensions}

The coefficient of the $g_{\mu\nu}$ term in the field equations can be 
adjusted through the
variation of a power of the determinant $g$ with respect to a symmetric 
metric.  In
particular, the variational derivative
$${{\delta (-g)^p}\over {\delta g_{\mu\nu}}}~=~2p~(-g)^p~g^{\mu\nu}
\eqno(24)$$
implies that
$$\delta~\int~d^D x~(-g)^p~(R-2 \Lambda)~=~\int d^D x~(-g)^p~[R_{\mu\nu}~-~
p R g_{\mu\nu}~+~2p \Lambda g_{\mu\nu}]~\delta g^{\mu\nu}
\eqno(25)$$
so that solutions of the equations
$$R_{\mu\nu}~-~p~R~g_{\mu\nu}~+~2p~\Lambda g_{\mu\nu}~=~0
\eqno(26)$$
would represent extrema of the modified action.

By considering diffeomorphisms $\delta g^{\mu\nu}~=~{1\over 2}(\xi^{\mu;\nu}~+~
\xi^{\nu;\mu})$, 
invariance of the modified action
can be verified only for those transformations which preserve the condition of
constant curvature, $\xi^\mu~{R,}_\mu~=~0$.   
In two dimensions, however, there always exists a conformal mapping from a 
metric on a Riemann surface to a constant curvature metric.  
Since an arbitrary diffeomorphism can be 
regarded as the product of a diffeomorphism of the type considered above and 
a conformal transformation, one might consider the existence of modified 
actions in two dimensions with conformal symmetry and restricted 
diffeomorphism invariance.

In particular, a function $f(x)$ may be defined to be $det(g_{\mu\nu})$ 
at every
point $x$ on the manifold.  Under a conformal transformation $g_{\mu\nu} \to
\tilde g_{\mu\nu}~=~e^{-2 \sigma}g_{\mu\nu}$, $f(x) \to \tilde f(x)~=~
e^{-2 \sigma} f(x)$.  Since 
$ln~f(x)$ is a scalar field, its covariant derivative is
an ordinary derivative rather than the special derivative associated with a
function of a scalar density $\sqrt g$.  Consequently, $\Delta_g ~ln~f(x) 
\not= 0$ and
$$\Delta_{\tilde g}~ln \tilde f(x)~=~e^{2 \sigma}~\Delta_{g}~ln~f(x)~-~
4~e^{2 \sigma} \Delta_g \sigma
\eqno(27)$$
so that $\int~d^2 \xi \sqrt{g}~(R~-~{1\over 4} \Delta_g~ln~f)$ is
invariant under conformal transformations.  Similarly,
$$\eqalign{\int~d^2 \xi~ {\tilde g}^p~(\tilde R~-~{1\over 4} 
\Delta_{\tilde g}~ln~f)^{2p}~=&~
\int~d^2 \xi~e^{-4 p \sigma}~g^p\cdot e^{4p \sigma}(R~-~{1\over 4} \Delta_g~
ln~f)^{2p}
\cr
~=&~ \int~d^2 \xi~g^p~(R~-~{1\over 4} \Delta_g~ln~f)^{2p}
\cr}
\eqno(28)$$

However, the restricted diffeomorphism invariance of the action $\int d^2 \xi
~g^p~R$ is broken by the additional term.  The variation of the action (28) is
$$\eqalign{\delta~\int~d^2 \xi~g^p~(R~-~{1\over 4} \Delta_g~ln~f)^{2p}~=~
p~\int~d^2 \xi~&g^p~[(R~-~{1\over 4} \Delta_g~ln~f)^{2p}]_{; \mu} \xi^\mu
\cr
~-~
{p\over 2}~\int&~d^2 \xi~g^p~
[(R~-~{1\over 4} \Delta_g~ln~f)^{2p-1}]^{; \nu}_{;\nu \mu} \xi^\mu
\cr
-~p~\int~& g^p~R_{; \mu}~\xi^\mu~[(R~-~{1\over 4}\Delta_g~ln~f)^{2p-1}]
\cr}
\eqno(29)$$
where use of the on-shell condition $R_{\mu\nu}~=~{1\over 2}~R~g_{\mu\nu}$ 
has been made. This expression vanishes for $p~=~{1\over 2}$, for which the 
integral (28) reduces to $\int~d^2 \xi \sqrt g R$, a topological invariant.   
While a conformal transformation can be found such that $R~-~{1\over 4}
\Delta_g~ln~f~=~constant$, the variation will then be non-vanishing in general
because R is not constant on this metric slice.
For a restricted
diffeomorphism satisfying $R_{; \mu} \xi^\mu~=~0$, the first term would only
vanish if $(\Delta_g~ln~f)_{; \mu} \xi^\mu~=~0$, which is a non-trivial 
constraint on 
$\xi^\mu$ 
since $\Delta_g~ln~f$ has a gradient vector pointing in 
a direction different from the gradient vector of $R$.  The vanishing of the
second term in the variation (29) would lead to additional conditions on 
$\xi^\mu$.

Other possible generalizations of two-dimensional gravity, which have been 
considered 
in the literature, include $f(R)$ theories [14][15]. 

\noindent{\bf 4. A Different Form of the Gravitational Action in Four 
Dimensions and Diffeomorphism Invariance}

In four dimensions, the modified actions $\int~d^4x~(-g)^p~R^q$ will yield
equations of motion generalizing equation (26) for $\Lambda~=~0$.  By adding 
two
flat dimensions to a two-dimensional manifold, and adapting the argument given
above to four-dimensional manifolds, the lack of invariance of 
$\int~d^4x~(-g)^p~R^{2p}$ under the transformation 
$\delta g_{\mu\nu}~=~\xi_{(\mu;\nu)}$ when $p~\not={1\over 2}$ can be 
demonstrated.  When $q~\not=~2p,~p~\not={1\over 2}$, the field
equations will contain $qR_{\mu\nu}~-~pRg_{\mu\nu}$, which is not proportional
to $R_{\mu\nu}~-~{1\over 2}R g_{\mu\nu}$, and not 
covariantly constant.  Thus, even under infinitesimal diffeomorphisms, 
$\int~d^4x~(-g)^p~R^q$  will only be invariant when $p={1\over 2}$.

The form of the field equation (26) suggests that the addition of actions with
different values of $p$ could yield a variation involving $(R_{\mu\nu}~-~
{1\over 2} R g_{\mu\nu}) \delta g^{\mu\nu}$ after averaging of the
coefficients of $R_{\mu\nu}$ and $g_{\mu\nu}$.  Since
$$(-g)^{y_m}~=~(-g)^{y_0}[1~+~(y_m~-~y_0)~ln~(-g)~+~{{(y_m~-~y_0)^2}\over 2}~
{(ln~(-g))^2}~+~...~]
\eqno(30)$$
the following equality holds for the variation
$$\eqalign{\delta~\biggl[~\int~d^4x& ~x_0~(-g)^{y_0}~(R~-~2\Lambda_0)~+~
\int~d^4x~x_1~
 (-g)^{y_1}~(R~-~2\Lambda_1)~+~...
\cr
&~~~~~+~\int~d^4 x~x_n~(-g)^{y_n}~(R~-~2\Lambda_n)~\biggr]
\cr
~=&~\int~d^4 x~(-g)^{y_0}~(x_0~R_{\mu\nu}~-~x_0~y_0~R~g_{\mu\nu}~+~2~
\Lambda_0~x_0
~y_0~g_{\mu\nu})~\delta g^{\mu\nu}
\cr
&~~~~~+~\int~d^4 x (-g)^{y_1}~(x_1~R_{\mu\nu}~-~
x_1~y_1~R~g_{\mu\nu}~+~2~x_1~y_1~\Lambda_1~g_{\mu\nu})~\delta g^{\mu\nu}
\cr
&~~+~...
~+~\int~d^4 x~(-g)^{y_n}~(x_n~R_{\mu\nu}~-~x_n~y_n~R~g_{\mu\nu}~+~2~x_n~y_n~
\Lambda_n~g_{\mu\nu})~\delta g^{\mu\nu}
\cr}$$
$$\eqalign{
~=&~\int~d^4 x~(-g)^{y_0}~[(x_0~+~x_1~+~
.~+~x_n)~R_{\mu\nu}
\cr
&~~~~~~~~~~~~~~~-~(x_0y_0~+~x_1y_1~+~...~+~x_ny_n)~R~g_{\mu\nu}
\cr
&~~~~~~~~~~~~~~~+~
2(x_0y_0\Lambda_0~+~x_1y_1\Lambda_1~+~...~+~x_ny_n\Lambda_n)~g_{\mu\nu}]~
\delta g^{\mu\nu}
\cr
&~~~~~+~\int~d^4 x (-g)^{y_0} ln (-g)[((y_1-y_0)x_1~+~...~+~
(y_n-y_0)x_n)~R_{\mu\nu}
\cr
&~~~~~~~~~~~~~~~~-~(x_1y_1(y_1-y_0)~+~...~+~x_ny_n(y_n-y_0))~R~
g_{\mu\nu}
\cr
&~~~~~~~~~~~~~~~~+~ 2(x_1 y_1 (y_1-y_0)\Lambda_1~+~...~+~x_ny_n 
(y_n-y_0)\Lambda_n)~g_{\mu\nu}]~\delta g^{\mu\nu}
\cr
&~~~~~+~...~
\cr}
\eqno(31)$$
By setting
$$\eqalign{x_0~+~x_1~+~...~+~x_n~=&~1
\cr
x_0y_0~+~x_1y_1~+~...~+~x_ny_n~=&~{1\over 2}
\cr
-x_1y_1(y_1~-~y_0)~-...-~x_ny_n(y_n~-~y_0)~=&~-{1\over 2}[~x_1(y_1~-~y_0)
~+...+~x_n(y_n~-~y_0)~]
\cr
&~=~-{1\over 2}({1\over 2}-y_0)
\cr
&\vdots
\cr}
\eqno(32)$$
including relations for the variables $\{x_0,~y_0,...,x_n,~y_n\}$ corresponding
to the higher order terms $(ln(-g))^m$, one obtains variations at each
order in the Taylor series expansion that are of the form
$$\eqalign{\int d^4 x(-g)^{y_0}(ln (-g))^m 
&
\left[{1\over {m!}}({1\over 2}-y_0)^m(R_{\mu\nu}-{1\over 2}
Rg_{\mu\nu})-\sum_{i=1}^n x_iy_i{{(y_i-y_0)^m}\over {m!}} \Lambda_i 
g_{\mu\nu}\right]\delta g^{\mu\nu}
\cr}
\eqno(33)$$
The conditions (32) can be reformulated as
$$\left(\matrix{1&1&1&1&...&1&
\cr
y_0&y_1&y_2&y_3&...&y_n&
\cr
y_0^2&y_1^2&y_2^2&y_3^2&...&y_n^2&
\cr
y_0^3&y_1^3&y_2^3&y_3^3&...&y_n^3&
\cr
y_0^4&y_1^4&y_2^4&y_3^4&...&y_n^4&
\cr
y_0^5&y_1^5&y_2^5&y_3^5&...&y_n^5&
\cr
\vdots&\vdots&\vdots&\vdots&\vdots&\vdots&
\cr}
\right)
\left(\matrix{x_0&
\cr
x_1&
\cr
x_2&
\cr
x_3&
\cr
&
\cr
x_n&
\cr}
\right)~=~\left(\matrix{1&
\cr
{1\over 2}&
\cr
{1\over 4}&
\cr
{1\over 8}&
\cr
{1\over {16}}&
\cr
{1\over {32}}&
\cr
\vdots &
\cr}
\right)
\eqno(34)$$
Given values of $y_0,~y_1,~y_2,~...,~y_n$, there will exist solutions to  
equation (34),
$x_0,~x_1,~x_2,...,$
$x_n$, only if the rank of the $n \times \infty$ matrix is at most $n+1$.
Besides the trivial solution $y_0~=~y_1~=~...~y_n~=~{1\over 2}$, the
generic solution to the set of simultaneous equations represented by (34)
involves an infinite number of coefficients $\{x_i\}$ and powers $\{y_i\}$.
The existence of a solution at finite $n$ can be determined as follows.
Suppose, for example, one considers the simplest non-trivial case
$y_0~=~y_1~=~...~=~y_{q-1}~=~y_{q+1}~=~...~=~y_n~\ne~y_q$ for some 
$0~\le~q~\le~n$.  Assuming that $n+1$ rows of the matrix, $(1,...,1)$,
$(y_0^{m_1},...,y_n^{m_1})$, 
\hfil\break
$...,(y_0^{m_n},...,y_n^{m_n})$ are linearly
independent, the decomposition of the row $(y_0^l,...,y_n^l)$ leads to the
relation $C_{l0}~+~C_{l1}y_0^{m_1}~+~...~+~C_{ln}y_0^{m_n}~=~y_0^l$
together with the subsidiary condition 
$C_{l0}~+~{1\over {2^{m_1}}}C_{l1}
~+~...~+~{1\over {2^{m_n}}}C_{ln}~=~{1\over {2^l}}$. Defining 
$\lambda_q~=~{{y_q}\over {y_0}}$,  the decomposition of the $q$th entry
gives $C_{l0}~+~\lambda_q^{m_1}y_0^{m_1}C_{l_1}~+~
~+~\lambda_q^{m_2}y_0^{m_2}C_{l2}~+~...~+~\lambda_q^{m_n}y_0^{m_n}C_{l_n}
~=~\lambda_q^l y_0^l$ which is an equation with $l$ complex solutions
and $l_R$ real solutions.  For each row labelled by $l$ and
corresponding set of $n+1$ coefficients $\{C_{li}\}$, there exists
$l_R$ real solutions for the variable $\lambda_q$. Thus, a necessary 
condition for the existence of a solution to equation (34) at finite $n$
is that the intersection of the common roots of the algebraic
equations for $\lambda_q$ consists of at least one real number.      
Finally, since each of the integrals is SL(4;R)-invariant,
the sum of the integrals in (31) will be invariant under only SL(4;R)
transformations unless further conditions on $x_i,~y_i$ and $\Lambda_i$
are imposed.

For an arbitrary infinitesimal diffeomorphism 
$\delta g^{\mu\nu}~=~{1\over 2}(\xi^{\mu;\nu}+\xi^{\nu;\mu})$,
$$\eqalign{\int~d^4x(-g)^{y_0}(ln(-g))^m&(R_{\mu\nu}-{1\over 2} g_{\mu\nu}) 
\xi^{\mu;
\nu}
\cr
&~=~\int~d^4x[(-g)^{y_0} (ln(-g))^m (R_{\mu\nu}-{1\over 2} R g_{\mu\nu})
\xi^\mu]^{,\nu}
\cr
&~~~~~-~2m~\int~d^4x (-g)^{y_0}(ln(-g))^{m-1} {\Gamma^\rho}_{\tau\rho}
 (R_\mu^\tau-{1\over 2} R \delta_\mu^\tau) \xi^\mu
\cr
&~~~~~+~(1-2y_0)~\int~d^4x (-g)^{y_0}(ln(-g))^m {\Gamma^\rho}_{\tau\rho}
(R_\mu^\tau-{1\over 2}R\delta_\mu^\tau) \xi^\mu
\cr} 
\eqno(35)$$
Addition of the terms (33) at each order leads to a cancellation of the
integrals containing ${\Gamma^\rho}_{\tau\rho}$ and leaves integrals of total 
derivatives which
vanish when $\xi^\mu$ tends to zero at spatial infinity.

Similarly,
$$\eqalign{ 2~\int~d^4x (-g)^{y_0} (ln(-g))^m&\left(\sum_{i=1}^\infty~x_iy_i
 {{(y_i-y_0)^m}\over {m!}}\Lambda_i \right)~\xi^\mu_{;\mu}
\cr
&~=~2 \sum_{i=1}^\infty~x_iy_i{{(y_i-y_0)^m}\over {m!}}\Lambda_i
\cdot \biggl[~\int~d^4x~[(-g)^{y_0}(ln(-g))^m \xi^\mu]_{,\mu}
        \cr
&~~~~~~~~~~~~~~~~~~~~~~~~~~~-~2m\int d^4x (-g)^{y_0} (ln(-g))^{m-1} 
                     {\Gamma^\rho}_{\mu\rho}\xi^\mu
                       \cr
&~~~~~~~~~~~~~~~~~~~~~+(1-2y_0)~\int ~ d^4x (-g)^{y_0} (ln(-g))^m 
                {\Gamma^\rho}_{\mu\rho} \xi^\mu \biggr]
                        \cr}
\eqno(36)$$
and cancellation of integrals containing ${\Gamma^\rho}_{\mu\rho}$ to all 
orders
occurs if
$$\eqalign{(1-2y_0)\sum_{i=0}^\infty~x_iy_i\Lambda_i~-~2
\sum_{i=1}~x_iy_i (y_i-y_0)
\Lambda_i~&=~0
\cr
(1-2y_0)\sum_{i=1}^\infty~x_iy_i {{(y_i-y_0)^m}\over {m!}} \Lambda_i
-2(m+1)~\sum_{i=1}^\infty~x_i y_i{{(y_i-y_0)^{m+1}}\over {(m+1)!}}\Lambda_i
~&=~0
\cr
m~&\ge 1
\cr}
\eqno(37)$$
>From the relations (37), it follows that all of the cosmological constant
terms in the field equations, resulting from the vanishing of the variations
(33) are equal.  The field equations then can be written
as a single equation 
$$R_{\mu\nu}~-~{1\over 2}~R~g_{\mu\nu}~+~2~\sum_{i=1}^\infty~x_i~y_i~\Lambda_i
g_{\mu\nu}~=~0
\eqno(38)$$

The field equations (38) also can be derived from a simplified action using the
identities (32).  The Ricci scalar terms may be combined since
$$\eqalign{\int~d^4x~&\sum_{i=0}^\infty~x_i~(-g)^{y_i}~R
\cr
~&=~\int~d^4x~\left[~x_0~
(-g)^{y_0}~R~+~\sum_{i=1}^\infty~x_i~R~(-g)^{y_0}~\sum_{m=1}^\infty
\left[{{(y_i-y_0)^m (ln(-g))^m}\over {m!}} \right]~\right]
\cr
~&=~\int~d^4x~\sum_{i=0}^\infty~x_i~(-g)^{y_0}~R
~+~\int~d^4x~(-g)^{y_0}~R~\sum_{m=1}^\infty
{{({1\over 2}-y_0)^m~(ln~(-g))^m}\over {m!}}
\cr
~&=~\int~d^4x~(-g)^{1\over 2}~R
\cr}
\eqno(39)$$
The cosmological constant terms can be combined because
$$\eqalign{\sum_{i=1}^\infty&~x_i (y_i-y_0)^{m+1} \Lambda_i~+~
y_0~\sum_{i=1}^\infty~x_i (y_i-y_0)^m \Lambda_i~=~\sum_{i=1}^\infty~
x_i y_i(y_i-y_0)^m \Lambda_i
\cr
\sum_{i=1}^\infty~x_i& y_i (y_i-y_0)^m \Lambda_i~=~({1\over 2}-y_0)~
\sum_{i=1}^\infty~x_i y_i (y_i-y_0)^{m-1} \Lambda_i~=~...
\cr
~&=~\left({1\over 2}-y_0\right)^{m-1}~\sum_{i=1}^\infty~x_i (y_i-y_0) \Lambda_i
~=~\left({1\over 2}-y_0\right)^m~\sum_{i=0}^\infty~x_i~y_i \Lambda_i
\cr}
\eqno(40)
$$
The solution to (40) is
$$\sum_{i=1}^\infty~x_i (y_i-y_0)^m \Lambda_i~=~2~\left({1\over 2}-y_0
\right)^m
~\sum_{i=0}^\infty~x_i y_i \Lambda_i
\eqno(41)$$
so that the integral is
$$\eqalign{2~&\int~d^4x~(-g)^{y_0}~\sum_{i=0}^\infty~x_i \Lambda_i
~+~2~\int~d^4x~(-g)^{y_0}~\sum_{m=1}^\infty
\sum_{i=1}^\infty~x_i (y_i-y_0)^m \Lambda_i~{{(ln(-g))^m}\over {m!}}
\cr
~&=~2~\int~d^4x~\sum_{i=0}^\infty~x_i \Lambda_i~+~
~4~\sum_{m=1}^\infty\int~d^4x~(-g)^{y_0} 
\left({1\over 2}-y_0\right)^m~{{(ln(-g))^m}\over {m!}}
\sum_{i=1}^\infty~x_i y_i \Lambda_i
\cr}
\eqno(42)$$
This gives
$$2 \int~d^4x~(-g)^{y_0}~\sum_{i=0}^\infty~x_i \Lambda_i~+~
4~\int~d^4x~(-g)^{1\over 2}~\sum_{i=1}^\infty~x_i y_i  \Lambda_i
~-~4~\int~d^4x~(-g)^{y_0}~\sum_{i=0}^\infty~x_i y_i \Lambda_i
\eqno(43)
$$
and the first and third terms cancel because
$$\sum_{i=0}^\infty~x_i (y_i-y_0) \Lambda_i~=~2 \left({1\over 2}-y_0 \right)
\sum_{i=0}^\infty~x_i y_i \Lambda_i
\eqno(44)
$$
so that the standard action
$$\int~d^4x~(-g)^{1\over 2}~(R~-~4~\sum_{i=1}^\infty~x_i y_i \Lambda_i)
\eqno(45)$$
is obtained.

\noindent{\bf 5. Gravitational Perturbations and Modifications of the Action}

The connection between the cosmological constant problem and general
covariance of the action has been noted in the
context of a different choice of gravitational action [16]-[18][19].  

Suppose that there exists a region of four-dimensional asymptotically flat 
space-time which represents a solution of the field equations of the standard 
Einstein-Hilbert action 
\hfil\break
$\int~d^4 x \sqrt{-g} R$ with zero cosmological constant.  A perturbation of 
this metric, through the addition of a finite
matter distribution in the space-time, for example, will lead
to a finite addition to the action.  The vanishing of the curvature in the 
asymptotically flat regions leads to finiteness
of the additional curvature contributions to the action in (31), whereas
the $\Lambda_i$ terms in this action appear to give rise to a divergent
contribution.  However, one could introduce theoretically a finite number of 
non-zero $\{\Lambda_m\}$ by choosing $x_ m \Lambda_m< = {1\over V}$, 
where $V$ is a regularized space-time volume.

Relaxing the 
requirement of infinitesimal diffeomorphism invariance temporarily, 
and allowing for a greater arbitrariness in the coefficients $x_i$ and powers
$y_i$,  suggests a possible description of localized perturbations [20] 
through an action of the type given in (31). 
Each of the $SL(4;R)$-invariant
integrals $I_i = \int d^4x (-g)^{y_i} (R - 2\Lambda_i)$ can be used to 
represent the action associated with a matter distribution localized in a 
region in space-time.  A generic matter distribution might curve space-time
in several regions, and it can be represented by $\sum_i~x_i~I_i$ for
an appropriate choice of $\{x_i\}$ since there are enough available
parameters to solve the relation $\sum_i x_i I_i = \delta I$, 
where $\delta I$ is the contribution of the perturbed metric to the 
gravitational action.  Previously, deviations in the metric resulting from
specified matter distributions have been described using gravitational 
actions containing either $R$ or $f(R)$,
and then determining the contributions to the action, to second order,
for example, in the scalar, vector and tensor perturbation variables 
by finding the variational principles giving rise to the
perturbed Einstein equations [21].  Cosmological constants were not
included in these variational principles, but the methods do not preclude
{\it a priori} the presence of a cosmological constant in the full
gravitational field equations, and indeed, solutions to such equations
could exist.  As an example, a special type of matter distribution giving
rise to a geometry with de Sitter metric inside a finite volume [22]
has been mentioned in a study of the definition of the localized energy density
for de Sitter space and perturbed metrics [23].
 The advantage of the approach used in this section is that
no {\it a priori} assumptions are made with respect to the magnitude
of the cosmological constant terms in actions representing inhomogeneous
metric perturbations, and instead, their vanishing in the large-volume limit 
shall be deduced.

In a universe with radius of curvature $10^{27}$ cm,  the observational limit
on the cosmological constant is $10^{-54}$ cm$^{-2}$.  Consider the 
Friedmann-Robertson-Walker metric
$$ds^2~=~-dt^2~+~R(t)^2\left[ {{dr^2}\over {1-k r^2}}~+~r^2(d\theta^2~+~
sin^2 \theta~d\phi^2)~\right]
\eqno(46)$$
with a scale factor $R(t)~\sim~t^{2\over 3}$ in the matter-dominated era
and $R(t)~\sim~t^{1\over 2}$ in the radiation-dominated era.  
When $k~=~0$,
$$\int~d^4x~{\sqrt {-g}}~=~\int~dt~dr~d\theta~d\phi~R^3(t)~r^2~sin~\theta
~=~{{4 \pi r^3}\over 3}~\int~dt~R^3(t) 
\eqno(47)$$
and
$$\eqalign{\int~d^4x~(-g)^{y_i}~&=~\int~d^4x~
[R^6(t)\cdot r^4 sin^2 \theta]^{y_i}
~=~\int~dt dr d\theta d\phi~R^{6y_i}(t)\cdot r^{4 y_i} sin^{2y_i} \theta
\cr
~&=~{{2 \pi r^{1+4y_i}}\over {1+4y_i}}~\int~dt~R^{6y_i}(t)
~\int_0^\pi~sin^{2y_i} \theta~d \theta
\cr}
\eqno(48)$$
it follows that
$$\int~d^4x~(-g)^{y_i}~\sim~r^{1+4y_i}~t^{1+4y_i}~\sim~V^{{1+4y_i}\over 3}
\eqno(49)$$
in the matter-dominated universe
and finiteness of the action in (31) leads to the conditions
$$x_i~\Lambda_i~\sim~{1\over {V^{{1+4y_i}\over 3}}}
\eqno(50)$$
whereas
$$V^{{1+3y_i}\over 3}~<~\int~d^4x~(-g)^{y_i}~\sim~r^{1+4y_i}~t^{1+3y_i}
~<~V^{{2(1+4 y_i)}\over 5}      
\eqno(51)$$
in the radiation-dominated universe, so that
$$x_i~\Lambda_i<{1\over {V^{{1+3y_i}\over 3}}}
\eqno(52)$$
Thus, if there any values of $y_i$ less than ${1\over 2}$, the conditions (50)
and (52) allow for the possibility of $x_i~\Lambda_i$ not decreasing as fast
as $1\over V$.    

Without any restriction on the coefficients $x_i$ and powers $y_i$,
arbitrary localized
\hfil\break
perturbations in the metric can be described by
a sum of integrals of the type 
\hfil\break
$\sum_{i=0}^\infty \int d^4x~x_i (-g)^{y_i}
(R-2\Lambda_i)$ with some of the values of $x_i~\Lambda_i$ possibly being 
greater than the observational limits on the cosmological constant.
The $SL(4;R)$ invariance of these integrals might be
regarded as sufficient to view the sum as a type of gravitational action [19].
The representations of localized perturbations by this sum allows for a
more coherent resolution of the problem of their contribution to the
overall cosmological constant, as it may be contrasted with
$\epsilon \int d^4x {\sqrt {-g}} (R~-~2\Lambda)$, $\epsilon~\ll~1$, which
cannot be directly used for their description.

Reimposing infinitesimal diffeomorphism invariance leads to the
conditions (32) and (37) on $x_i$ and $y_i$, which imply that the effective
cosmological constant contribution to the action is
$-4~\int~d^4x~(-g)^{1\over 2}~\sum_{i=0}^\infty~x_i y_i \Lambda_i$.
Finiteness of this integral requires that 
$\sum_i~x_i y_i \Lambda_i~\sim~{1\over V}$, and there are likely to be
cancellations amongst the coefficients $x_i$ corresponding to those powers 
$\{y_i\vert y_i~<~{1\over 2}\}$, so that the observational limits will again
be satisfied by the terms $x_i~\Lambda_i$ for each i.  Because of the
restrictions on $x_i$ and $y_i$ resulting from infinitesimal diffeomorphism
invariance, the action in (31) is no longer relevant in the description of
localized perturbations.  However, the overall effect of the finite action
condition is sufficient to ensure that the large-scale cosmological constant
should be set to zero even in the modified version of the gravitational action.

Non-zero values of $\Lambda_i,~i=0,1,2,...$ are feasible over a 
limited region of the four-
\hfil\break
dimensional space-time (with the remaining volume being described by a zero
cosmological constant) or for a closed universe, because
a divergence in the action is prevented by the integration of a convergent 
sum $\sum_{i=0}^\infty~x_i y_i \Lambda_i$ over a finite four-volume.    
The inflationary universe [24][25] provides the standard example of the 
introduction of a non-zero cosmological constant over a limited four-volume.   

Gravitational perturbations have successfully been described by higher-order 
curvature terms [21][26].  They have even been used 
to explain the density spectrum [27].  Nonlinear gravitational Lagrangians 
give rise to field equations which often have solutions that approximate 
solutions of the Einstein field equations and standard metrics such as that of 
de Sitter space have been shown to arise as an attractor in the space of 
solutions of higher-derivative gravity theories [28]. 
 
\noindent{\bf 6. f(R) Theories and other Generalizations}

Another type of generalized gravitational action which leads to the Einstein 
field equations
with cosmological constant is characterized by a function of the Ricci 
scalar [29][30]
\hfil\break
[31][32].
Viewing the Ricci scalar as a function of the independent variables, 
$g^{\mu\nu}$,
the metric, and $\Gamma^\rho_{\mu\nu}$, the connection,          
$R~=~R_{\mu\nu}(\Gamma) g^{\mu\nu}$, the variation of the action with respect 
to $g^{\mu\nu}$
gives
$${\delta \over {\delta g^{\mu\nu}}}~\int~d^4x~{\sqrt{-g}}~f(R)=~
\int~d^4x~{\sqrt{-g}} f^\prime(R) {{\delta R}\over {\delta g^{\mu\nu}}}~+~
\int~d^4x~{{\delta{\sqrt {-g}}}\over {\delta g^{\mu\nu}}}~f(R)
\eqno(53)$$
so that the field equations are 
$$f^\prime(R)~R_{\mu\nu}(\Gamma)~-~{1\over 2}~f(R)~g_{\mu\nu}~=~0
\eqno(54)$$
which has trace
$$f^\prime (R)~R~-~2~f(R)~=~0
\eqno(55)$$
If $R~=~c_i$ is a zero of this equation, then
$$c_i~=~2{{f(R)}\over {f^\prime(R)}}~\biggl\vert_{R=c_i}
\eqno(56)$$
and
$$R_{\mu\nu}~-~{1\over 4}~c_i~g_{\mu\nu}~=~0
\eqno(57)$$
the gravitational field equations with $\Lambda~=~{1\over 4}~c_i$.  

This technique can also be extended to functions of the Ricci tensor and 
Riemann
curvature tensor.  For example, if $S~=~g^{\mu\alpha}~g^{\nu\alpha}
~R_{\alpha\beta}(\Gamma)~R_{\mu\nu}(\Gamma)$ and $L(g,\Gamma)~=~f(S)$ [33], 
then the
field equations are
$$2~f^\prime(S)~R_{\alpha(\mu}R_{\nu)}{}^\alpha~-~{1\over 2}~f(S)~
g_{\mu\nu}~=~0
\eqno(58)$$
The trace of this equation is
$$f^\prime(S)~S~-~f(S)~=~0
\eqno(59)$$
which has roots $c_i^\prime~=~{{f(c_i^\prime)}\over {f^\prime(c_i^\prime)}}$.  
Substituting these values of S into (58) gives
$$R_{\alpha(\mu}~R_{\nu)}{}^\alpha~-~{1\over 4}~{c_i^\prime}~g_{\mu\nu}~=~0
\eqno(60)$$
Solutions of the equation
$$R_{\mu\nu}~-~{{\sqrt {c_i^\prime}}\over 2}~g_{\mu\nu}~=~0
\eqno(61)$$
will also be solutions of (58).

The square of the Riemann curvature tensor can be used to construct a
third type of action from which the gravitational field equations can be
derived. Let 
\hfil\break
${\hat S}~=~g_{\mu\alpha}~g^{\nu\beta}~g^{\rho\gamma}~g^{\lambda\delta}
R^\mu{}_{\nu\rho\lambda}(\Gamma)~R^\alpha{}_{\beta\gamma\delta}(\Gamma)$
$${\delta \over {\delta g^{\tau \sigma}}}~\int~d^4x~{\sqrt {-g}}~f({\hat S})
~=~\int~d^4x~{\sqrt {-g}}~f^\prime({\hat S})~{{\delta {\hat S}}
\over {\delta g^{\tau\sigma}}}
~-~\int~d^4x~{1\over 2}~f({\hat S})~g_{\mu\nu}
\eqno(62)$$
Since

$$\eqalign{
{{\delta {\hat S}}\over {\delta~g^{\tau\sigma}}}~=&~
-R_{(\tau}{}^{\nu\rho\lambda}
~R_{\sigma)\nu\rho\lambda}~+~R_{\alpha(\tau}{}^{\rho\lambda}~
R^\alpha{}_{\sigma)\rho\lambda}
~+~R_{\mu\nu(\tau}{}^\lambda~R^{\mu\nu}{}_{\sigma)}{}^\lambda~
\cr
~&+~R_{\mu\nu\rho(\tau}~R_{\sigma)}{}^{\mu\nu\rho}
\cr}
\eqno(63)$$
the field equations are
$$2~R_{(\tau}{}^{\nu\rho\lambda}~R_{\sigma)\nu\rho\lambda}~f^\prime({\hat S})
~-~{1\over 2}~g_{\tau\sigma}~f({\hat S})
~=~0
\eqno(64)
$$
which has trace
$$f^\prime({\hat S})~{\hat S}~-~f({\hat S})~=~0
\eqno(65)$$
The zeroes of this equation $c_i^{\prime\prime}~=~{{f(c_i^{\prime\prime})}
\over {f^\prime(c_i^{\prime\prime})}}$ and the Riemann tensor satisfies
$$R_{(\tau}{}^{\nu\rho\lambda}~R_{\sigma)\nu\rho\lambda}~-~
{1\over 4}{c_i^{\prime\prime}}~g_{\tau\sigma}~=~0
\eqno(66)$$
For a space-time with vanishing Weyl tensor, solutions of the field equations
$$R_{\mu\nu}~-~{{\sqrt {3c_i^{\prime\prime}}}\over 4}~g_{\mu\nu}
~=~0
\eqno(67)$$
will satisfy (64).

\noindent{\bf 7. Higher-Order Curvature Terms in Superstring 
Effective Actions} 

The same method can be applied to higher-derivative actions obtained in 
string theory.  Graviton amplitudes in superstring theory can be used to
deduce an effective action, which receives higher-order corrections at the 
three-loop level.  At third-order in the inverse string 
tension [34][35],
$$I~=~{1\over {2\kappa^2}}~\int~d^{10}x~{\sqrt {-g}}~\left(R~+~
{{\zeta(3)}\over {256}}
~\alpha^{\prime 3}~{\tilde Y}~\right)
\eqno(68)
$$
where
$$\eqalign{
{\tilde Y}~&=~{1\over 2}~(R_{abcd}~R^{abcd})^2~+~R^{abcd}~
R_{cdef}R^{efhg}~R_{hgab}
~+~8~R^{abcd}~R_{bedf}~R^{egfh}~R_{gahc}
\cr
~&~~~~~+~R^{abfg}~R_{bd}{}^{\!hk}~
R^{de}{}_{\!gh}~
R_{eakf}
~-~8~R_{abef}~R^{abfg}~R_{cdgh}~R^{cdhe}
\cr
~&~~~~~-~R_{abef}~R^{ab}{}_{\!gh}~R_{cd}{}^{\!fg}~R^{cdhe}
\cr}
\eqno(69)$$
By subtracting a term proportional to the dimensionally continued Euler form
$L_{(4)}$ [36][37]
\hfil\break
[38], a different
combination of curvature tensors can be used in the action 
$$I~=~{1\over {2 \kappa^2}}~\int~d^{10}x~{\sqrt {-g}}~\left(R~+~
{{\zeta(3)}\over {16}}
\alpha^{\prime 3}~Y\right)
\eqno(70)$$
with
$$Y~=~{1\over {16}}({\tilde Y}~-~{1\over {96}}{\L_{(4)}})
~=~2~R_{abcd}~R_e{}^{bc}{}_{\!f}~R^{aghe}~R^f{}_{gh}{}^d
~+~R_{abcd}~R_{ef}{}^{cd}~R^{aghe}~R^f{}_{gh}{}^{\!b}
\eqno(71)$$
The Euler-Lagrange equations of motion in the first-order formalism are then
$$R_{\tau\sigma}~-~{1\over 2}~R~g_{\tau\sigma}~
+~{{\zeta(3)\alpha^{\prime 3}}\over {16}}~\left[{{\delta Y}
\over {\delta g^{\tau\sigma}}}
~-~{1\over 2}~Y~g_{\tau\sigma}~\right]
~=~0
\eqno(72)$$
where
$$\eqalign{
{{\delta Y}\over {\delta g^{\tau\sigma}}}~=&~2R_{a(\tau\vert cd} 
R_{e\vert \sigma)}{}^c{}_{\!f}~R^{aghe}~R^f{}_{gh}{}^d
\cr
~+&~2R_{ab(\tau\vert d}~R_{e\vert}{}^b{}_{\!\sigma)
f}~R^{aghe}~R^b{}_{gh}{}^{\!d}
~+~2R_{abcd}~R_e{}^{bc}{}_f~R^{ag}{}_{(\tau|}{}^e~R^f{}_{g\vert \sigma)}{}^d
\cr
~+&~2R^a_{b(\tau\vert d}~R^e{}_{f\vert \sigma)}{}^d~R_a{}^{ghe}~
R^f{}_{gh}{}^b
~+~2R_{abcd}R_{ef}{}^{cd}~R^a{}_{(\tau}{}^{he}~R^f{}_{\sigma)h}{}^b
\cr}
\eqno(73)$$
The trace of equation (72) is
$$4R~+~{{\zeta(3)\alpha^{\prime 3}}\over {16}}~Y~=~0
\eqno(74)$$
When the Weyl tensor vanishes, and $R_{\mu\nu}~=~9~\kappa~g_{\mu\nu}$,
then $Y~=~4320~\kappa^4$ and equation (74) implies that
$$\kappa~=~-\left({4\over {3\zeta(3)}}\right)^{1\over 3}
{1\over {\alpha^\prime}}
\eqno(75)$$
and
$$R_{\mu\nu}~=~-\left({4\over {3 \zeta(3)}}\right)^{1\over 3}
~{9\over {\alpha^\prime}}~g_{\mu\nu}
\eqno(76)$$
so that the effective cosmological constant is 
$-\left({4\over {3 \zeta(3)}}\right)^{1\over 3}~{9\over {\alpha^\prime}}$.  
Higher orders in the expansion also produce contributions of order 
${1\over {\alpha^\prime}}$.

A conformal transformation of the maximally symmetric solution to (76), 
anti-de Sitter space with cosmological constant $-\left({4\over
{3 \zeta(3)}}\right)^{1\over 3} {9\over {\alpha^\prime}}$, to flat space 
exists, and redefinitions of 
the metric
at higher orders have been shown to be sufficient to maintain the validity of
the Ricci-flat condition $R_{\mu\nu}~=~0$ when the target space is
a Calabi-Yau manifold. 

Since redefinitions of fields in a Lagrangian do not change the S-matrix, 
transformations of the metric, anti-symmetric field and dilaton have been
used to write the effective bosonic string action up to $O(\alpha^{\prime 2})$
[39][40][41][42], and similar techniques can be used for the type IIB 
superstring and heterotic string.  At $O(\alpha^\prime)$ for the bosonic 
string, the 4-graviton amplitude requires the existence of the 
$R_{\mu\nu\rho\sigma}~
R^{\mu\nu\rho\sigma}$ term, but the $R^2$ and $R_{\mu\nu}~R^{\mu\nu}$ 
terms can be removed by a metric redefinition.  It is preferable to choose
the Gauss-Bonnet invariant 
$R_{\mu\nu\rho\sigma} R^{\mu\nu\rho\sigma} -4 R_{\mu\nu} R^{\mu\nu}
+ R^2$ so that ghosts do not arise upon quantization [36]. 
The Lovelock Lagrangians represent the only combinations of Riemann tensors
which give rise to field equations containing at most second-order derivatives
of the metric, and they are necessary for the absence of ghosts at this order.
At higher orders, there is a form of the effective action, with 
a minimal number of terms, such that additional terms besides
the dimensionally continued Lovelock invariants [43] seem to arise.  
These calculations have also not been performed beyond the order at which 
the Lovelock Lagrangians are no longer relevant.  For example, the quintic 
Lovelock tensor has only recently been evaluated [44].  The higher-order 
Lovelock Lagrangian
$$L_{(n)}~=~{{(2n)!}\over {2^n}}~R_{[i_1 i_2}{}^{i_1i_2}~R_{i_3i_4}{}^{i_3i_4}
~R_{i_5i_6}{}^{i_5i_6}~...~R_{i_{2n-1}i_{2n}]}{}^{i_{2n-1}i_{2n}}
\eqno(77)$$
only vanishes in ten dimensions when $n \ge 6$, and it is zero in twenty-six 
dimensions when $n \ge 14$.    

The elimination of ghosts is conventionally achieved by removing all
higher-order derivatives in the linearized equation for the spin-2 graviton
field $h_{\mu\nu}$, so that the $h~\nabla^{2n}~h$, $n~\ge~2$, 
terms are cancelled using the dimensionally continued Gauss-Bonnet terms 
[36][45][46] and the graviton propagator is not modified by extra powers 
of the momenta.  The interaction terms arise at $O(h^n),~n\ge 3$ in the 
expansion of the effective action.  It has been established that there 
exists a field redefinition that removes the higher order derivatives 
$h~\nabla^{2n}~h$, $n~\ge~2$, from the effective action [46] but does not 
affect the interaction terms.  Since a more arbitrary form could be assumed 
for the interactions, other types of higher-derivative curvature 
terms are, in principle, allowed by the conditions placed on the 
linearized field equations, and they do appear in string effective actions.  
Besides considering the linearized equations,
the field equations for the total metric $g_{\mu\nu}$ are of interest.
The inclusion of terms other than the Lovelock invariants leads to differential
equations with third-order and higher-order derivatives of $g_{\mu\nu}$.
If one wishes to obtain a field equation with at most second-order
derivatives of the total metric $g_{\mu\nu}$, then the Lovelock invariants
are required.       

It remains to establish through a detailed calculation which additional 
higher-order curvature terms can be removed through further metric 
redefinitions. For example, it might be possible to adapt the technique used
for removal of quadratic terms in $h_{\mu\nu}$ [46] to cubic and higher-order
terms.  If it is feasible to use repeated application of  
appropriate metric redefinitions to eliminate higher-order derivatives 
at $O(h^n)$, $n \ge 3$, then there may exist truncation of the series 
expansion in the effective action.   

Such a result has been obtained in special cases.
It is well-known, for example, that the Ricci-flat condition can be 
maintained through a metric redefinition when the target space is a Calabi-Yau 
manifold [47][48], with the proof making use of the property that the 
manifold is Kahler.  Moreover, several results have been obtained for
higher-derivative theories.  When the Lagrangian is an arbitrary function 
of the Ricci scalar, there exists a conformal transformation from the 
higher-derivative theory to the Einstein-Hilbert action with a scalar 
field [28][29][49], and Legendre transformations 
[50][51] have been used to transform Lagrangians dependent on the 
Ricci scalar and Ricci tensor to the Einstein-Hilbert action coupled to 
scalar fields and fields of up to spin 2.

Restriction to the Lovelock invariants leads to consideration of Lagrangians
of the type $f(L_{(n)})$ in the first-order formalism.  Defining
$G_{(n)}^{\mu\nu}$ to be the variation of
${\sqrt {-g}}~L_{(n)}$ with respect to $g_{\mu\nu}$ [52], it follows that
$$G_{(n) \mu\nu}~=~{-1\over {2^{n+1}}}g_{\mu\sigma}
\delta^{\sigma i_1 ... i_{2n}}_{\nu j_1 ... j_{2n}}~R^{j_1j_2}{}_{i_1 i_2}
~...R^{j_{2n-1}j_{2n}}{}_{i_{2n-1} i_{2n}}
\eqno(78)$$
and, specifically,
$$\eqalign{G_{(0)\mu\nu}~&=~-{1\over 2}~g_{\mu\nu} 
\cr
G_{(1) \mu\nu}~&=~R_{\mu\nu}~-~{1\over 2}~R~g_{\mu\nu}
\cr
G_{(2)\mu\nu}~&=~2(R_\mu{}^{\rho\sigma\lambda}~R_{\nu\rho\sigma\lambda}
~-~2~R_{\mu\rho\nu\lambda}~R^{\lambda\rho}~-~2~R_{\mu\rho}~R^\rho_\nu
~+~R~R_{\mu\nu})
\cr
&~~~~~-~{1\over 2} (R^{\mu\nu\rho\lambda}~R_{\mu\nu\rho\lambda}
~-~4~R^{\mu\nu}~R_{\mu\nu}~+~R^2)~g_{\mu\nu}
\cr
~& \vdots
\cr}
\eqno(79)$$
If the degree of the function $f$ is $r$, then the field equations
$$f^\prime(L_{(n)})~{{\delta~L_{(n)}}\over {\delta~g_{\mu\nu}}}
~-~{1\over 2}~g^{\mu\nu}~f(L_{(n)})~=~0
\eqno(80)$$
are differential equations of degree $2r$, but they will consist of factors
which contain at most second-order derivatives of the metric.
The differential equations (80) can be solved by simultaneously
satisfying the following conditions 
$$\eqalign{{{\delta~L_{(n)}}\over {\delta~g_{\mu\nu}}}~&=~0
\cr
L_{(n)}~&=~0
\cr}
\eqno(81)$$
which both represent second-order differential equations in the metric.  
More generally, contracting 
equation (80) gives 
$$f^\prime(L_{(n)})~g^{\mu\nu}~{{\partial L_{(n)}}\over {\partial g_{\mu\nu}}}
~-~{D\over 2}~f(L_{(n)})~=~0
\eqno(82)$$
where $D$ is the space-time dimension.  From (79), it follows that
$g^{\mu\nu}~{{\partial L_{(n)}}\over {\partial g_{\mu\nu}}}
~=~d_{(n)}L_{(n)}$ for some constant $d_{(n)}$.  If $L_{(n)}~=~c_{(n)}$
is a root of
$$d_{(n)}~f^\prime(L_{(n)})~L_{(n)}~-~{D\over 2}~f(L_{(n)})~=~0
\eqno(83)$$
then
$$c_{(n)}~=~{D\over {2d_{(n)}}}~{{f(c_{(n)})}\over {f^\prime(c_{(n)})}}
\eqno(84)$$
and the equation of motion (80) becomes
$${{\delta L_{(n)}}\over {\delta g_{\mu\nu}}}~-~c_{(n)}~d_{(n)}~g^{\mu\nu}
~=~0
\eqno(85)$$
 
Gravitational perturbations which are local and not homogeneous, giving rise to
an energy-momentum tensor not proportional to the metric, have already been 
shown to be best described by $SL(4;R)$-invariant gravitational action terms 
such as those considered in $\S 5$ and higher-derivative curvature terms 
[21][26].  Moreover, it has been shown in $\S 6$ that the Einstein field 
equations with cosmological constant can be derived 
from $f(R)$ theories and other generalizations.  Finally, redefinition of the 
dilaton field in the four-dimensional string effective action is known to 
change the value of the cosmological constant [53].  

It follows that the matter distributions at both Planck scales and 
macroscopic scales can be represented adequately by an action which 
contains the Einstein-Hilbert Lagrangian plus higher-order curvature terms, 
such as those that arise in the string effective actions, and 
other matter Lagrangians, ultimately derived from the string theory.  The 
cosmological constant should be regarded 
as a flexible variable, which can be shifted to zero using symmetry
transformations, rather than as a fixed parameter. 

The solutions to the resulting field equations will, in general, be 
represented by phases 
with both $\Lambda~=~0$ and $\Lambda~\ne~0$, and it may seem 
that the different values of $\Lambda$ are equally probable.  
However, it has been shown that the field equations
have special properties for the value $\Lambda~=~0$.  Specifically, all of the 
generalized symmetries, based on prolongations of vector fields in the space 
of coordinates and metrics, have been classified for the four-dimensional 
vacuum field equations with both $\Lambda~=~0$ and $\Lambda~\ne~0$ 
\hfil\break
[54][55].  When $\Lambda~=~0$ and 
$h_{\mu\nu}(x^a, g_{ab},g_{ab,c_1},...,g_{ab,c_1,...c_k})$ are the components 
of a k-th order 
generalized symmetry of the vacuum equations, there is a constant C and 
a generalized 
vector field $X^i~=~X^i(x^a,g_{ab}, g_{ab,c_1},...,
g_{ab,c_1,...,c_{k-1}})$ such that, modulo the field equations, 
$h_{ab}~=~C~g_{ab}~+~\nabla_a~X_b~+~\nabla_b~X_a$ [54].  When 
$\Lambda~\ne~0$, the uniform scale
transformations do not leave the vacuum equations invariant.

This result suggests a mechanism for explaining the vanishing of 
$\Lambda$ at classical scales.
Given different phases with $\Lambda~=~0$ and $\Lambda~\ne~0$,  a conformal 
transformation 
representing a time-dependent expansion, similar to the scale factor of the 
Friedmann-Robertson-Walker universe, would increase the size of the 
$\Lambda~=~0$ region 
relative to the sizes of the $\Lambda~\ne~0$ regions.  Over the course of 
the cosmological expansion, the dominance of the $\Lambda~=~0$ region becomes 
apparent.  From the above considerations, this mechanism is applicable to 
general matter distributions and effective actions containing a variety 
of different higher-derivative terms. 

This mechanism is similar to the irreversible processes in the thermodynamics
of many-body systems [56][57][58], which result from
reversible processes at the microscopic level.  The relation between
microscopic and macroscopic physics has been considered in the context of
irreversible kinetic equations [58] and the Onsager repricocity relations
[56].  These techniques in non-equilibrium statistical mechanics
have been adapted to general relativity [59] and the cosmological constant 
problem using the fluctuation-dissipation theorem [60][61].  Since
microscopic fluctuations are accompanied by dissipative processes in
macroscopic systems, it has been suggested, for example, 
that dissipative particle creation in de Sitter space, represented 
by the non-symmetric part of the polarization tensor, leads to the
diminishing of the effective cosmological constant [60].   

A variant of this approach to macroscopic physics is required in the present
model.   Conformal transformations and field redefinitions which
leave the S-matrix invariant but shift $\Lambda$ in the effective action [53]
can be viewed as microscopic reversible transformations
$\Lambda_1~\to~\Lambda_2$, $\Lambda_2~\to~\Lambda_3$ ... $\Lambda_{N-1}~\to~
\Lambda_N$, $\Lambda_N~\to~\Lambda_1$, $N~\ge 2$.  

The field equations at macroscopic levels will be modifications of 
the Einstein field equations. Inhomogeneous localized matter distributions 
are not relevant for the cosmological constant term, and from the generalized
symmetries of the vacuum field equations, it will be shown that the scale
invariance of the $\Lambda~=~0$ equations leads to an irreversible
process mapping regions with cosmological constant $\Lambda_i\ne 0$ to
regions with $\Lambda~=~0$. More generally, from mesoscopic to large scales, 
the global asymptotics of $T_{\mu\nu}$ should be compatible with dilations of 
the three-dimensional hypersurface metric, if it is assumed that the 
energy-momentum tensor is derived from 
a conformally invariant string theory and tends to the perfect-fluid tensor
in the large-volume limit.  The cosmological solutions to the field equations 
obviously reflect this scale invariance, and, in particular, the spatial
sections of the Friedmann-Robertson-Walker universe increase by a scale factor 
$R^2(t)$, while the time coordinate can be rescaled so that the entire
four-dimensional metric is uniformly dilated.

The reversible microscopic transformations initially will lead to
infinitesimal conformal transformations of the three-dimensional hypersurface
metric $~^{(3)}g_{\alpha\beta}~\to~(1~+~\epsilon)~^{(3)}g_{\alpha\beta}$.
The rescaling of the tetrad $e_{0t}~\to~e_{0t},~e_{it}~\to~e_{it},~
e_{0\alpha}~\to~(1+\epsilon)^{1\over 2} e_{0\alpha},~e_{i\alpha}~\to~
(1+\epsilon)^{1\over 2}e_{i\alpha}$ determines the transformations of 
the spin connection $\omega_{\mu m n}~=~{1\over 2}e_m^\nu(\partial_\mu 
e_{n \nu}~-~\partial_\nu e_{n \mu})$~
\hfil\break
$-~{1\over 2} e_n^\nu(\partial_\mu e_{m\nu}
~-~\partial_\nu e_{m \mu})~-~{1\over 2}e_m^\rho e_n^\sigma (\partial_\rho
e_{c \sigma}~-~\partial_\sigma e_{c \rho})e^c_\mu$ and the Ricci tensor
\hfil\break
$R_{\nu\sigma}~=~e^{m \mu} e^n_\sigma (\partial_\mu \omega_{\nu m n}
~-~\partial_\nu \omega_{\mu m n}~+~\omega_{\mu m}{}^c \omega_{\nu c n}
~-~\omega_{\nu m}{}^c \omega_{\mu c n})$. The transformation of the
field equations, restricted to the three-dimensional hypersurface, can then
be either calculated directly or deduced by noting that
$~^{(3)}R_{\alpha\beta}~\to~~^{(3)}R_{\alpha\beta}$ while the conjugate
momenta to the metric, $\pi^{\alpha\beta}~=~{{\delta I}\over 
{\delta~^{(3)}{\dot g}_{\alpha\beta}}}$, change to $(1+\epsilon)^{-1} 
\pi^{\alpha\beta}$, since the three-dimensional scale invariance is 
assumed to be a property of the effective action at mesoscopic to large 
scales.  Transposing any conjugate momenta 
terms to the other side of the equation, one finds that a zero cosmological 
constant in the constraint equations will be maintained, while the 
a non-zero cosmological constant will be altered by an infinitesimal amount
since $\Lambda~^{(3)}g_{\alpha\beta}~\to~(1+\epsilon)~\Lambda~^{(3)}
g_{\alpha\beta}$.  Thus, $\Lambda~=~0$ remains a fixed point in the
infinitesimal time evolution of three-surface metrics in cosmological
space-times, while the $\Lambda~\ne~0$ sectors will be shifted by conformal
transformations, and also by other field redefinitions.   
           
Conformal transformations will dilate the
regions with $\Lambda~=~0$ and their existence in the macroscopic order,
formed during the cosmological evolution,
can be ensured by a fluctuation-enhancement theorem [57].  In this scheme, 
the order-parameter would define the lack of randomness in the choice of the
value of $\Lambda$ in the effective field equations.
   
This explanation for the vanishing of the cosmological constant may be 
contrasted with several recently proposed solutions using higher-dimensional 
gravity and duality in string theory.  It has recently been shown that 
general relativity
with a cosmological constant appears as an effective theory at the 
four-dimensional boundary of a Chern-Simons Lagrangian integrated over 
five-dimensional manifold.  When the gauge group is SO(1,5), the action is
$$\eqalign{I_{CS}~&=~{1\over 2}~\int_{{\cal M}_4 \times R}~
\epsilon_{ABCDE}~\biggl[{\tilde R}^{AB}~\wedge~{\tilde R}^{CD}~\wedge~
e^E
~-~{{2 \sigma}\over {3L^2}} {\tilde R}^{AB}~\wedge~e^C~\wedge~e^D~\wedge~e^E
\cr
&~~~~~~~~~~~~~~~~~~~~~~~~~~~~~~~~~~~~~~+~{1\over
{5L^4}}~e^A~\wedge~e^B~\wedge~e^C~\wedge~e^D~\wedge~e^E
\biggr]
\cr
{\tilde R}^{AB}~&=~d \omega^{AB}~+~\omega^A{}_C~\wedge~\omega^{CB}
\cr}
\eqno(86)$$
where $\sigma~=~\pm 1$, L is an arbitrary length parameter, ${\cal M}_4$ is 
four-dimensional manifold with boundary $\partial {\cal M}_3$, $e^A$ is 
vielbein and $\omega^{AB}$ is the five-dimensional spin connection [62].     
By requiring the variation of the action to be zero at the boundary, 
and imposing conditions on the components of the vielbein and spin connection
normal to the boundary, one obtains the standard 
four-dimensional gravitational field equations with cosmological constant
$$\Lambda~=~{1 \over {l^2}}~+~{\sigma\over {3L^2}}
\eqno(87)$$
where $l$ is another length parameter used in the definition of the  
normal components of the vielbein and spin connection.  The value of
$\Lambda$ can be set to zero,  although one may note the special 
choice of parameters, $\sigma~=~-1$ and $l^2~=~3L^2$, 
the absence of a matter distribution and the
derivation of the field equations from a classical action.
          
It is also of interest to note that Chern-Simons Lagrangians in odd dimensions
with gauge group $SO(1,2n-1)$ can be decomposed with respect to the
group $SO(1, 2n-2)$ [63],
$$I_{CS}~=~2n~\sum_{k=0}^{n-1}~\left({{n-1}\atop k}\right)~[2(n-1-k)]!~
(-1)^{n-1-k}
l^{-(2n-1-2k)}~L_{(k)}~+~total~der.
\eqno(88)$$
so that in 5 dimensions, the sum consists of a cosmological constant term,
a Ricci scalar and the quadratic Gauss-Bonnet invariant.  From the variations  
in the first-order formalism in $\S 6$, it has been established that the
higher-derivative terms introduce another cosmological constant term into
the field equations for a particular set of solutions, confirming the
expression (87) for $\Lambda$ as a sum of two different contributions.
It also follows from (88) that if the Lovelock invariants $L_{(0)}$,..., 
$L_{(4)}$ that appear in the 10-dimensional superstring effective action      
occur in the linear combination 
$$10\left[~8!~l^{-9} L_{(0)}~-~4\cdot 6!~l^{-7}~L_{(1)}
~+~6\cdot 4!~l^{-5}~L_{(2)}~-~4\cdot 2!~l^{-3}~L_{(3)}~+~l^{-1}~L_{(4)}
~\right]
\eqno(89)$$ 
they will describe a Chern-Simons Lagrangian with gauge group SO(1,9).

Target-space duality has been used to relate solutions to three-dimensional
string equations of motion described by space-times with $\Lambda~\ne~0$
and $\Lambda~=~0$ [64][65].  Breaking of a discrete
symmetry such as T-duality may lead to a strict partitioning of the
space of induced worldsheet metrics implying the breaking of
conformal invariance, and a mechanism for inducing such symmetry breaking has 
yet to be developed. S-duality invariance of the field equations derived
from a four-dimensional superstring effective action has also been studied
[66][67].  The duality invariance of the field equations when $\Lambda~=~0$, 
and non-invariance when $\Lambda~\ne~0$, mirrors the
conformal invariance of the gravitational field equations discussed earlier.
However, since S-duality is a discrete symmetry, again it may not be
as useful as conformal invariance in obtaining a proof of the vanishing of the
cosmological constant. 

Matter fields often give rise to an effective $\Lambda$, and fine-tuning  of 
the
parameters in a cosmological model is necessary to obtain vanishing $\Lambda$ 
[8].   A slight change in the parameters often leads to the re-appearance of 
the cosmological constant [7].  The path integral approach has been used 
previously to justify vanishing of $\Lambda$, and it is dependent 
on the choice of
action determining the weighting of the metrics and an approximation in the 
summation over space-time topologies [68][69].  The Einstein-Hilbert action 
that is conventionally used in the path integral can be modified by 
additional curvature terms which appear in the superstring effective action.  
The higher-derivative terms that typically arise in these effective actions 
lead to an alternative gravitational description of the perturbations 
associated with the inclusion of matter fields.  It is being suggested here 
that the symmetries in string theory, especially conformal invariance, 
provide an additional way of eliminating the cosmological 
constant at macroscopic scales.  They are also likely to be relevant 
in a complete study of the cosmological constant problem at the quantum 
level, where other techniques have been developed [70][71][72].

The classical version of the cosmological constant problem considered in 
this paper concerns the fitting of space-time metrics, which solve $\Lambda=0$
field equations, with realistic matter distributions.  Within the context
of a generalized gravity theory, presumably derived from string theory,
a variational argument for the vanishing of $\Lambda$ has been put forward by 
applying specific symmetry transformations to the action and field equations
and then establishing that there is an irreversible flow to the $\Lambda=0$
sector.
\vfill
\eject
\centerline {\bf Acknowledgements.}  
\vskip 5pt
\noindent
I would like to thank Prof. D. G. Crighton, Dr. G. W. Gibbons and 
Prof. S. W. Hawking for their encouragement.  Useful discussions with 
Dr F. Embacher, Dr M. Perry and Dr A. Tseytlin about higher-order curvature
terms and path integrals in quantum gravity are also gratefully acknowledged.
The calculations of the symmetric and non-symmetric metric variations of the
action in $\S 2$ were done in 1993 under the auspices of the Research
Foundation of Southern California, while the actions in two and four 
dimensions in $\S 3$ and $\S 4$, invariant under restricted coordinate
transformations, were constructed in the first half of 1994 at Cambridge.   
\vfill
\eject
\centerline{\bf REFERENCES}
\item{[1]}  M. Ferraris, M. Francaviglia and I. Volovich, Il Niovo Cim. 
${\underline{108B}}$
(1993) 1313 - 1317
\item{[2]}  V. H. Hamity and D. E. Barraco, Gen. Rel. Grav. 
${\underline{25}}$ (1993) 461 - 471
\item{[3]}  G. Magnano and L. M. Sokolowski, Phys. Rev. ${\underline{D50}}$ 
(1994) 5039 - 5059
\item{[4]}  R. Capovilla, Nucl. Phys. ${\underline{B373}}$ (1992) 233 - 246
\item{[5]}  I. Bengtsson and P. Peldan, Phys. Lett. ${\underline{B244}}$ 
(1990) 261 - 264
\hfil\break
I. Bengtsson, Phys. Lett. ${\underline{B254}}$ (1991) 55 - 60
\hfil\break
I. Bengtsson and O. Bostrom, Class. Quantum Grav. ${\underline 9}$ 
(1992) L47 - L51
\item{[6]}  P. Peldan, Class. Quantum Grav. ${\underline{11}}$ (1994) 1087 - 
1132
\item{[7]}  S. Weinberg, Rev. Mod. Phys. ${\underline{61}}$ (1989) 1 - 23
\item{[8]}  S. M. Carroll, W. H. Press and E. L. Turner, Annual Rev. Astron. 
Astrophys.
${\underline{30}}$ (1992) 499 - 542
\item{[9]}  A. Einstein and E. G. Straus, Ann. Math. ${\underline{47}}$
(1946) 731 - 741
\item{[10]}  J. W. Moffat, Phys. Rev. ${\underline{D 19}}$ (1979) 3554 - 3558
\item{[11]} T. Damour, S, Deser and J. McCarthy, Phys. Rev. 
${\underline {D45}}$ (1992) 3289 - 3291 ; Phys. Rev. ${\underline{D47}}$ (1993)
1541 - 1556    
\item{[12]}  C. W. Misner, K. S. Thorne, J. A. Wheeler, 
${\underline{Gravitation}}$
(New York: Freeman,
\hfil\break
1973)
\item{[13]}  C. Moller,  ${\underline{The~Theory~of~Relativity}}$ 
(Oxford:Clarendon
 Press, 1952)
\item{[14]}  S. Mignemi and H.-J. Schmidt, Class. Quantum Grav. 
${\underline{12}}$ (1995) 849 - 857
\item{[15]}  S. N. Solodukhin, Phys. Rev. ${\underline{D51}}$ 
(1995) 591 - 602
\item{[16]}  W. G. Unruh, Phys. Rev. ${\underline{D40}}$ (1989) 1048 - 1052 
\hfil\break
 M. Henneaux and C. Teitelboim, Phys. Lett. $\underline{B222}$ 
(1989) 195-199
\item{[17]}  M. Kreuzer, Class. Quantum Grav. ${\underline 7}$ (1990) 1303 - 
1317
\item{[18]}  Y. J. Ng and H. van Dam, Phys. Rev. Lett. ${\underline{65}}$
(1990) 
1972 -1974; J. Math. Phys. ${\underline{32}}$ (1991) 1337 -1340
\item{[19]}  A. Zee, `Remarks on the Cosmological Constant Problem'
Proc. of 20th Annual Orbis Scientiae, dedicated to Dirac's 80th year, 
Miami, Fla., Jan. 17 - 22, 1983, ed. by S. L. Mintz and A. Perlmutter
(New York: Plenum, 1985) 211 - 230
\item{[20]}  N. Deruelle, J. Katz and J.-P. Uzan, `Integral Constraints
on Cosmological Perturbations and their Energy', gr-qc/9608046
\item{[21]}  V. F. Mukhanov, H. A. Feldman and R. H. Brandenberger,
Phys. Rep. ${\underline{215C}}$ (1992) 203 - 233
\item{[22]}  A. Einstein, Sitzungsberichte Der K${\ddot o}$niglisch 
Preussischen Akademie 
\hfil\break
Der Wissenschaften (1918), No. XII, 270 - 272
\item{[23]}  J. W. Maluf, `On the Distribution of Gravitational Energy
in De Sitter Space', gr-qc/9608051
\item{[24]}  A. H. Guth, Phys. Rev. ${\underline{D23}}$ (1981) 347 - 356
\hfil\break
 A. Linde, Phys. Lett. ${\underline{B108}}$ (1982) 389 - 393
\hfil\break
A. Albrecht and P. J. Steinhardt, Phys. Rev. Lett. ${\underline{48}}$ (1982) 
1220 - 1223
\item{[25]}  A. Linde, ${\underline{Particle~Physics~and~Inflationary~
Cosmology}}$,
 Contemporary Concepts in
\hfil\break
Physics, Vol. 5 (Chur: Harwood, 1990)
\item{[26]}  D. S. Salopek and J. Stewart, Phys. Rev. ${\underline{D51}}$
(1995) 517 - 535
\item{[27]}  A. Lukas, Phys. Lett. ${\underline{B347}}$ (1995) 13 - 20
\item{[28]}  A. B. Mayer and H.-J. Schmidt,  Class. Quantum Grav.
${\underline{10}}$ (1993) 2441 - 2446 
\item{[29]}  J. D. Barrow and A. Ottewill, J. Phys. A ${\underline{16}}$ 
(1983) 2757 - 2776
\hfil\break
J. D. Barrow and S. Cotsakis, Phys. Lett. ${\underline{B214}}$ (1988) 515 - 518
\item{[30]}  K. Maeda, Phys. Rev. ${\underline{D39}}$ (1989) 3159 - 3162
\item{[31]}  A. Borowiec, M. Ferraris, M. Francaviglia and I. Volovich, Gen. 
Rel. Grav.
${\underline{26}}$ (1994) 637 - 644
\hfil\break  
I. V. Volovich, Mod. Phys. Lett. A ${\underline{8}}$ (1993) 1827 - 1834
\item{[32]}  S. Rippl, H. van Elst, R. Tavakol and D. Taylor, `Kinematics and 
Dynamics of f(R) Theories of Gravity', gr-qc/9511010
\item{[33]}  A. Borowiec, M. Ferraris, M. Francaviglia and I. V. Volovich,
`New Lagrangians for Einstein Equations and Signature Change', A.93, in
\hfil\break
${\underline{Preliminary~Programme,~Extended~Abstracts~of~Plenary}}$
\hfil\break
${\underline{Lectures~and~Abstracts~of~Contributed~Papers}}$, 14th GR 
International Conference
on General Relativity and Gravitation, Florence, Italy, 6 - 12 
August 1995        
\item{[34]}  D. J. Gross and E. Witten, Nucl. Phys. 
${\underline{B277}}$ (1986) 1 - 10
\item{[35]}  R. C. Myers, Nucl. Phys. ${\underline{B289}}$ (1987) 701 - 716
\item{[36]}  B. Zwiebach, Phys. Lett. ${\underline{B156}}$ (1985) 315 - 317   
\item{[37]}  F. M\"uller-Hoissen, Class. Quantum Grav. ${\underline{3}}$ (1986)
665 - 667
\item{[38]}  F. M\"uller-Hoissen, Nucl. Phys. ${\underline{B337}}$ (1990)
709 - 736
\item{[39]}  M. C. Bento and N. E. Mavromatos, Phys. Lett. 
${\underline{B190}}$ (1987) 105 - 109
\item{[40]}  A. A. Tseytlin, Phys. Lett. ${\underline{B176}}$ (1986) 92 - 98;
Nucl. Phys. ${\underline{B294}}$ (1987) 383 - 411
\item{[41]}  R. R. Matsaev and A. A. Tseytlin, Phys. Lett. 
${\underline{B185}}$ (1987) 52 - 58; Phys. Lett. ${\underline{B191}}$ (1987)
354 - 362
\item{[42]}  M. C. Bento and O. Bertolami, Phys. Lett. ${\underline{B228}}$
(1989) 348 - 354
\item{[43]}  D. R. T. Jones and A. M. Lawrence, Z. Phys. C - Particles and 
Fields ${\underline{42}}$ (1989) 153 - 158
\item{[44]}  C. C. Briggs, `A General Expression for the Quintic Lovelock 
Tensor', gr-qc/9607033
\item{[45]}  S. Deser and A. N. Redlich, Phys. Lett. ${\underline{B176}}$
(1986) 350 - 354
\hfil\break
S. Deser, Physica Scripta, Vol. T15 (1987) 138 - 142
\item{[46]}  I. Jack, D. R. T. Jones and A. M. Lawrence, Phys. Lett. 
${\underline {B203}}$ (1988) 378 - 380
\item{[47]}  E. Witten, Nucl. Phys. ${\underline{B268}}$ (1986) 79 - 112
\item{[48]}  D. Nemeschansky and A. Sen, Phys. Lett. ${\underline{B178}}$
(1986) 365 - 369
\item{[49]}  G. Magnano, M. Ferraris and M. Francaviglia, Gen. Rel. Grav.
${\underline{19}}$ (1987) 465 - 479
\hfil\break
H.-J. Schmidt, Astron. Nachr. ${\underline{308}}$ (1987) 183 - 188
\hfil\break
A. Jakubiec and J. Kijowski, Phys. Rev. ${\underline{D37}}$ (1988) 
1406 - 1409
\item{[50]}  M. Ferraris, M. Francaviglia and G. Magnano, Class. Quantum 
Grav. ${\underline 5}$
(1988) L95 - L99
\item{[51]}  G. Magnano, M. Ferraris and M. Francaviglia, Class. Quantum 
Grav. ${\underline 7}$
(1990) 557 - 570; J. Math. Phys. ${\underline{31}}$(2) (1990) 378 - 387
\item{[52]}  F. M\"uller-Hoissen, Phys. Lett. ${\underline{B163}}$ (1985) 
106 - 110
\item{[53]}  N. Kaloper, Phys. Lett. ${\underline{B336}}$ (1994) 11 - 17
\item{[54]}  I. M. Anderson and C. G. Torre, Commun. Math. Phys. 
${\underline{176}}$ (3) (1996) 479 - 539
\item{[55]}  L. Marchildon, `Lie Symmetries of Einstein's Vacuum Equations
in N Dimensions', A.119, in
${\underline{Preliminary~Programme,~Extended~Abstracts~of~Plenary~  
Lectures}}$
\hfil\break
${\underline{~and~Abstracts~of~Contributed~Papers}}$,
14th GR International Conference on
 General Relativity and Gravitation, 
Florence, Italy, 6 - 12 August 1995
\item{[56]} L. S. Garcia-Colin and J. L. Del Rio, Rev. Mexicana de Fisica,
${\underline{39}}$ (1993) 669 - 684 
\item{[57]} M. Suzuki, `Theory of Instability, Nonlinear Brownian Motion and
Formation of
\hfil\break
 Macroscopic', ${\underline {Order~and~Fluctuations~in~Equilibrium ~and~
Non-Equilibrium}}$
\hfil\break
${\underline{~Statistical~Mechanics}}$ ed. by 
G. Nicolis, G. Dewel and J. W. Turner
(New York: John Wiley, 1981) 299 - 366
\item{[58]} P. Resibos, `Entropy and Irreversibility',
\hfil\break
 ${\underline{Non-Equilibrium~
Thermodynamics,~Variational~Techniques~and~Stability}}$, 
\hfil\break
ed. by R. J. Donnelly, R. Herman
and I. Prigogine (Chicago: University of Chicago Press, 1966) 239 - 248 
\item{[59]}  G. Ludwig, `Irreversibilit${\ddot a}$t und Kosmologie',
${\underline{Entstehung,~Enticklung~und~}}$
\hfil\break
${\underline{Perspektiven~der~Einsteinschen~
Gravitationstheories}}$ Deutsche Akademie 
der Wissenschaften zu Berlin (Berlin: Akademie-Verlag, 1966) 279 - 299
\item{[60]}  E. Mottola, Phys. Rev. ${\underline{D33}}$ (1986) 2136 - 2146
\item{[61]}  E. Mottola, `Fluctuation-Dissipation Theorem in General 
Relativity and
the Cosmological Constant' ${\underline {The~Physical~Origins~of~Time~ 
Asymmetry}}$
ed. by J. J. Halliwell, J. Perez-Mercader and W. H. Zurek (Cambridge: 
Cambridge University
Press, 1994) 504 - 515
\item{[62]}  M. Banados, `General Relativity from five dimensional 
Chern-Simons theory', 
\hfil\break
gr-qc/9603029
\item{[63]} F. M\"uller-Hoissen, Annalen der Physik, ${\underline 7}$
(1991) 543 - 557
\item{[64]}  G. Horowitz and D. L. Welsh, Phys. Rev. Lett. 
${\underline{71}}$ (1993) 328 - 331
\item{[65]} E. Alvarez, L. Alvarez-Gaume, J. F. L. Barbon and
Y. Lozano, Nucl. Phys. ${\underline{B415}}$ (1994) 71 - 100
\item{[66]} J. Maharana and H. Singh, Phys. Lett. ${\underline{B368}}$
(1996) 64 - 70
\item{[67]}  S. Kar, J. Maharana and H. Singh, Phys. Lett.
${\underline{B374}}$ (1996) 43 - 48
\item{[68]}  S. W. Hawking, Nucl Phys. ${\underline{B144}}$ (1978) 349 - 362
\item{[69]}  S. Coleman, Nucl. Phys. ${\underline{B310}}$ (1988) 643 - 685
\item{[70]}  G. Moore, Nucl. Phys. ${\underline{B293}}$ (1987) 139 - 188
\item{[71]}  K. R. Dienes, Phys. Rev. ${\underline{D42}}$ (1990) 
2004 - 2021
\item{[72]}  E. Witten, Int. J. Mod. Phys. ${\underline{A10}}$ (1995)
1247 - 1248

\end